\documentstyle[prd,preprint,aps]{revtex}

\newcommand{\be}{\begin{equation}}
\newcommand{\ee}{\end{equation}}
\newcommand{\bea}{\begin{eqnarray}}
\newcommand{\eea}{\end{eqnarray}}

\newcommand{\V} {{\cal V}}

\newcommand{\rx} {{\hat\rho}_r}

\begin{document}

\reversemarginpar
\tighten

\title{Anatomy of a Bounce} 

\author {A.J.M. Medved}

\address{
Department of Physics and Theoretical Physics Institute\\
University of Alberta\\
Edmonton, Canada T6G-2J1\\
E-Mail: amedved@phys.ualberta.ca\\}

\maketitle

\begin{abstract}

Holographic considerations are  used in the  scrutiny of
 a special  class of brane-world cosmologies.
Inherently to this class,  the brane 
 typically bounces, at a finite size, as a consequence 
of  a  charged black hole in the bulk.
Whereas a prior treatment \cite{BBB} 
emphasized 
 a brane that is void of  standard-model matter, 
the analysis is now extended to include
 an intrinsic (radiation-dominated) matter source. 
  An interesting  feature of this
generalized  model is that a bounce is no longer guaranteed
but, rather, depends on the  initial conditions. 
Ultimately, we demonstrate that compliance 
with an appropriate holographic bound
is  a sufficient prerequisite for  a  bounce to occur.

\end{abstract}

\section{Introduction}

In recent years, 
there has been dramatic progress in our understanding
of  cosmology. Nonetheless, many questions remain
unanswered and this will, in all likelihood,
continue to be the case until a fundamental  theory
of quantum gravity can be realized. (For some
relevant discussion, see \cite{SMO}.)
For instance, the very early stages of the universe,
where {\it transplanckian} effects must certainly be
accounted for, can not yet  properly be addressed.

One of the more notorious issues of the transplanckian universe
is that, from at least a  classical perspective, a big bang/crunch
 type of singularity seems to
be an inevitable feature of most cosmological models. 
It is often  hoped that some (yet unknown)
effects of quantum gravity can smooth out this
singular behavior. An alternative means of
circumvention  is the notion of 
a universe that smoothly  ``bounces'' (rather than collapsing) 
at some finite value of
the cosmological scale factor. Recent directions along this line 
have often employed a  brane-world setting \cite{RUB}: 
colliding branes can be used to effectively describe  the 
so-called  cyclic  theories
({\it e.g.}, \cite{KOSST}),  and  a  bouncing brane world
can be induced by, for instance, 
a suitably compactified string-theory background
\cite{BKM}, more than one large extra dimension \cite{BRAX} and
an electrostatic charge in the bulk \cite{MP}. (For
 a more generic discussion on bounce cosmologies
and earlier references, see \cite{BLAH}.)

In spite of the obvious virtues of a bounce cosmology,
there is (at least) one significant concern that must
be addressed  in such scenarios. Namely, bounce cosmologies are generally
 not possible  unless  the {\it null energy condition}  
(which, given causality,
requires  non-negative energy densities \cite{WALD})
 has somewhere been violated 
\cite{BLAH}.\footnote{We say ``generally'' because this need not be the case 
for a universe with closed spacelike slices.}
Hence, the pertinent question becomes whether such a violation should
 necessarily be regarded as a catastrophic event.
On one hand, 
violations of the null condition  
are known to be commonplace at the quantum level
and can even occur in purely classical circumstances
\cite{BV}. On the other hand, it is pretty much accepted that
there must be some modest limit to the degree of violation ({\it e.g.},
\cite{FORD}),
although it is often quite difficult to express
this limit in quantitative terms.

In view of the above discussion, an important
issue becomes the discrimination of {\it exotic} 
cosmologies ({\it i.e.}, cosmologies  containing negative-energy matter).
That is to say, we require, ideally speaking,
some means of testing the validity of an exotic 
spacetime, without resorting to the rather dubious  energy conditions
of general relativity \cite{BV}.
Notably, the  holographic paradigm \cite{THO,SUS,BOU2} 
can serve (and has served) nicely
in the proposed manner \cite{BFM,MCI2}. To briefly elaborate, the
holographic principle  provides a natural
bound on the amount of information and, hence, entropy
that can be stored in a given region of spacetime (When this
``holographic limit'' is saturated, a black hole can
be expected to form.)
It stands to reason that any cosmology that leads to a holographic violation
 (either directly by exceeding an upper bound on
the entropy or indirectly by contradicting a manifestation
of the principle)
  can {\it not}  be viewed as a realistic one.

Although the holographic principle is, in a sense, conceptually quite simple,
its application in a given context can often be  far from 
straightforward. This has  especially been  the case in
cosmological situations, where there has been considerable
debate as to how the principle should best be 
utilized \cite{FS,VEN,EL1,EL2,EL3} 
(also, \cite{BOU2} and references therein).
Such matters are further complicated for exotic cosmologies,
 as  it is clear  (intuitively speaking)  that all entropy bounds
will inevitably be threatened by negative-enough energies.
A formulation  of the principle that is especially well-suited
for generic (including  exotic!) cosmological scenarios 
 is  the {\it causal entropy bound}
\cite{CAU,BFV,BFM}. We will elaborate on the logistics
of this bound at an appropriate juncture in the paper.

Not long ago, the current author used the causal entropy
bound to test the holographic viability of a certain
class of exotic  cosmologies \cite{BBB}. 
More specifically, brane-world cosmologies ({\it a la} 
Randall and Sundrum \cite{RS}) such that
 a four-dimensional  brane  
 moves  through  the   five-dimensional anti-de Sitter  background
of a static  and  charged 
 (``Reissner-Nordstrom-like'')  black hole.\footnote{For
other studies on brane cosmologies with a charged black hole
bulk, see \cite{BC1,NEW,BC2,BC3,BC4,BC6,BC7,WOO,MP,AJM,YYY,YNOT,KTB}.}
The effective   cosmology of the brane (which is regarded
as ``our universe'') can be shown to translate 
into a Friedmann-Robertson-Walker (FRW) universe
with {\it holographically induced} matter.\footnote{This matter
is ``holographic'' in the sense that it is essentially a projection
of the bulk geometry onto the lower-dimensional  brane.
This notion of holography should not be confused with
the holographic principle {\it per se}.}
Along with a vacuum energy, this
 induced matter includes 
both  radiation   and exotic stiff matter; with
the latter being directly related to the bulk charge
and  responsible for
the occurrence of a  non-singular bounce. 
It was  ultimately shown that such cosmologies are indeed
holographically viable ({\it i.e.}, satisfy the casual
entropy bound at the bounce\footnote{The bounce hypersurface - that is,
the spacelike surface for which  the cosmological scale factor  reaches its 
minimal size -
is precisely where such a cosmology would be most susceptible
to a holographic violation.  This point will
be clarified later in the paper.})  provided that the bulk charge satisfies
a {\it lower} bound. This is, perhaps, a counter-intuitive outcome but
can be attributed to the exotic energy density (as calculated at the bounce) 
going as an inverse power of the charge.

In  the prior work of note \cite{BBB}, 
the brane was, for the sake of simplicity, regarded
as being empty {\it modulo}  the holographically induced matter.
The current paper will, however, extend considerations
to the more realistic case of a brane that also contains
an  intrinsic matter source; presumably, the contributions from the
standard model. This may seem, at first glance, to be a somewhat trivial
extension.  Nevertheless, it  is well known (see, {\it e.g.}, \cite{BCL})
that
such an inclusion leads to an unorthodox  energy-density-squared
term in the brane-world  cosmological equation.
As an immediate consequence of this ``addition'', there is now 
a distinct possibility that the brane universe no longer bounces.
(Note that  a bounce is an  inevitable 
feature of the empty-brane version, as long as the
bulk charge is non-vanishing \cite{MP,AJM}.)
In fact, it has been recently shown that, for a radiation-dominated
brane universe (which is also the interest of the current paper),
the bounce can only occur if the bulk charge satisfies a
finite lower bound \cite{KTB}.  We will eventually be able to show
 that this particular bound has a clear
 holographic interpretation.

Before proceeding, we should point out that a   recent paper
\cite{KTB} (also see \cite{MP}) has also considered this
 same  brane-world model  when it contains standard-model
(radiation-dominated) matter. Nonetheless, much of the emphasis  has been, 
up to now, 
on the phenomenological viability, whereas the
current treatment stresses  the holographic implications.
For further discussion on holography in the context of
brane worlds, see \cite{ZZZZZ}.

The rest of the paper is organized as follows. The next 
section discusses the forementioned brane-world cosmologies,
with special  attention paid  to the  solution
near the bounce (which is the most interesting
region from a holographic perspective).
In Section III, after a brief explanation
of the causal entropy bound, we
consider the implications of this holographic bound on
the cosmologies of interest. 
The final section contains a summary.

\section{Brane-World Cosmologies with Radiative Matter}

As in our prior related work \cite{BBB}, 
the model of interest is a certain class of
brane-world cosmologies 
for which a non-singular bounce is known to occur \cite{MP,AJM}.
However, unlike in the preceding study, we will now
incorporate the effects  of  intrinsic brane matter; that is,  
a ``non-holographic'' source that lives strictly on the brane 
({\it i.e.}, standard-model matter).
This inclusion is, of course, necessary for a  physically
realistic treatment of the problem.
It should, however, be kept in mind that the presence of intrinsic
brane matter can potentially  jeopardize
the existence of the bounce.

To be  precise, let us consider the scenario
of a 3+1-dimensional brane (possibly but not 
necessarily curved\footnote{Empirical
evidence suggests that our  universe and, hence, the brane should be
positively curved \cite{CAR}.
Nonetheless, the distinction between a flat and  ``de Sitter'' brane
 is inconsequential to
the regime of current  interest; namely, near the bounce.})
moving in an (otherwise static) 
4+1-dimensional anti-de Sitter bulk spacetime. Without loss of generality,
the geometry of this bulk  can be described by an
anti-de Sitter black hole with a constant-curvature horizon
and  an electrostatic charge \cite{NEW}. (Such
black holes may be regarded as Reissner-Nordstrom-like
but having an arbitrary horizon topology \cite{BIR}.)
It is, in particular, the presence of a non-vanishing
charge that can induce the desirable feature of a
 non-singular bounce.
We will therefore assume, for the duration,
that a charge is always present\footnote{Even if such a 
  black hole
is classically neutral, there are compelling reasons to believe
that quantum fluctuations would still  induce an effective
charge of significant magnitude \cite{GMX}.} and,
consequently, a bounce will generally occur (and will {\it always} occur
if the brane is empty).

Further commentary on  these  bouncing brane cosmologies
can be found in \cite{MP,AJM,YYY,BBB,KTB}. 
Typically, after a suitable tuning of the (effective) brane
cosmological constant, one obtains a universe
that, far from the bounce, tends asymptotically toward de Sitter space. 
However, as our
current interest is near the bounce,  such details,
although important in their own right,
are not of relevance to the current analysis. 

As is well documented, this type of  moving-brane scenario  leads
to an effective cosmology that resembles
a FRW universe  \cite{KRA}. (Keep in mind
that an observer on the brane perceives the motion through the bulk 
 as either a cosmological expansion or contraction.) 
In general terms, it is the nature
of the bulk solution that determines  what
types of matter will be {\it holographically induced} 
on the brane world. Furthermore, the current analysis will, as
alluded to above, also allow for matter that lives 
strictly on the brane.

We will now briefly outline the steps that lead up
to the cosmological equation of motion. More rigorous
discussions  can be found in ({\it e.g.}) \cite{KRA,PS,AJM2,ZZZZZ,AJM,YYY}
and references therein.

It is, first of all, necessary to 
explicitly formalize  the bulk solution;
in this case, a five-dimensional anti-de Sitter
spacetime with a  Reissner-Nordstrom-like  geometry. That 
is, 
\be
ds^2_{5}=-f(r)d\tau^2+{1\over f(r)}dr^2+r^2d\Omega^2_{k,3}\;,
\label{3.1}
\ee
where
\be
f(r)= {r^2\over L^2}+ k -{\omega M\over r^{2}}
+{3\omega^2 Q^2\over 16 r^4}\; ,
\label{3.2}
\ee
\be
\omega\equiv {16\pi G_5\over 3 \V_3} \; ,
\label{3.3}
\ee
and where  the following definitions  have  been employed:  
$L$ is the curvature radius of the anti-de
Sitter bulk, 
$G_5$ is the five-dimensional Newton constant
(which will be related to the usual four-dimensional constant, $G_4$, 
later on\footnote{Note that, here and throughout, all fundamental
constants, besides $G_4$ and $G_5$, are set equal to unity.})
and  $\V_3$ is the dimensionless
volume element associated with 
$d\Omega^2_{k,3}$
(a three-dimensional spacelike  hypersurface of constant curvature). 
 Also in evidence are
 three constants of integration, $k$, $M$ and $Q$,  which have the
following interpretations:
{\it (i)} $k$  is a discrete parameter that  describes the horizon topology, 
whereby $k=$  +1, 0 and -1 for
a spherical, flat and hyperbolic geometry (respectively), 
{\it (ii)} $M$ is the conserved mass  of the bulk black hole
and, for our purposes,  can be regarded as 
a strictly positive  quantity (however, see \cite{BIR}),
{\it (iii)}  $Q$ is the electrostatic charge of the black hole,
which - assuming  the existence of a pair of positive
and real horizons ({\it i.e.}, assuming cosmic censorship) - 
must be bounded in accordance with $Q^2<4 M^2/3\;$.
For this reason, we will usually work with
the following dimensionless measure of  charge:
\be
\epsilon^2 \equiv 3 Q^2 /4 M^2 <1\;. 
\label{3.4}
\ee

The next topic, on the agenda,  is the dynamical behavior of the
 brane world. After some suitable identifications,
it can be shown  that the induced metric on the
brane takes on  a FRW form,
\be
ds^2_{4}=-dt^2+a^2(t)d\Omega^2_{k,3}\;,
\label{3.5}
\ee
where $t$  is the physical time as measured by
 a brane observer and  
$a\equiv r(t)$ is the cosmological   scale factor.
Note that a dot will always denote 
a differentiation with respect to $t$.
 
After  further manipulations,  the corresponding Friedmann-like equation
 of motion is   found to be as follows:
\be
H^2= -{1\over L^2}-{k\over a^2}+{\omega M\over a^4}
-{\omega^2 M^2 \epsilon^2\over 4 a^6}+  
\left({8\pi G_5\over 3}\right)^2 \rho_{br}^2\;,
\label{3.6}
\ee
where $H\equiv {\dot a}/a$ is the Hubble parameter (constant to some)
and $\rho_{br}$ is the total  energy density of 
any  matter living  on the brane,
 including the  contribution from the brane tension.
(Note that $\rho_{br}$ enters the equation
of motion by application of the well-known  Israel junction conditions 
\cite{ISR}).

It is convenient to make a clear distinction between
the vacuum-energy contribution from the tension and the
``conventional'' matter sources on the brane. Hence, let us write
$\rho_{br}=\rho_0+\rho_{m}\;$, where $\rho_0$ is
the vacuum energy density and $\rho_{m}$  is the remainder.
It follows that the cosmological equation (\ref{3.6})
can be recast  in the following manner:
\be
H^2= {\Lambda_4\over 3}-{k\over a^2} + 
{8\pi G_4\over 3} \rho_{m} +
{\omega M\over a^4}
-{\omega^2 M^2 \epsilon^2\over 4 a^6}+  
{4\pi G_4\over 3 \rho_0} \rho_m^2\;,
\label{3.66}
\ee
where we have identified an {\it effective} cosmological
constant on the brane,
\be
\Lambda_4\equiv 3\left[\left({8\pi G_5\over 3}\right)^2\rho_0^2-{1\over L^2}
\right]\;,
\label{3.67}
\ee
and used standard expectations to relate the
bulk and brane-world gravitational  constants,
\be
{8\pi G_4\over 3}\equiv 2\left({8\pi G_5\over 3}\right)^2 \rho_0\;.
\label{3.68}
\ee
Note that, for the very special case of a  flat or {\it critical}  brane, 
one must finely tune $\Lambda_4$ to be a vanishing quantity.
Otherwise, in  general, one obtains a non-critical brane.

Remarkably, Eq.(\ref{3.66}) is, up to three extra terms,  
just the standard (four-dimensional)
Friedmann equation for arbitrary matter and  
curvature.\footnote{Note that this brane universe
can be open, closed or flat (respectively, $k=-1$, +1 or 0) depending on the 
horizon topology of the bulk solution.}
Meanwhile, the three extra terms include a holographic
radiative-matter contribution (the $a^{-4}$ term),
 a holographic  stiff-matter source
(the $a^{-6}$ term) and a quadratic energy-density term.
The quadratic term, although quite unorthodox, happens to be
a typical feature of many  brane-world cosmologies
\cite{BCL}. Let us also take  note of the 
negative-energy stiff matter.
It is this {\it exotic}   source - which is a direct manifestation 
of the bulk charge  - that can create a significant 
enough repulsive force (when the universe is small)
so that a singular  collapse may  be avoided.

Let us now, for the sake of definiteness,
 specialize to brane matter that is dominated
by radiation; hence, $\rho_m \sim a^{-4}\;$.
The reason for this choice is twofold.
Firstly, from a phenomenological standpoint, the early stages of the
observable universe are  certainly 
radiation dominated; suggesting that  radiation   
should similarly  be the dominant form in a near-bounce regime. 
 Secondly, in view of our holographic interest,  
radiation is clearly the most entropic form
of conventional  matter and, therefore,
would provide  the sternest test for any proposed
entropy bound. 

With the above in mind, let us adopt the notation 
\be
\rho_m=\rho_r={\rx\over a^4}
\label{3.69}
\ee
and then appropriately re-express the cosmological
equation (\ref{3.66}) as follows:
\be
H^2= {\Lambda_4\over 3}-{k\over a^2} + 
\left[ {8\pi G_4\over 3}\rx  + \omega M        \right]      
{1\over a^4}
-{\omega^2 M^2 \epsilon^2\over 4 a^6}+  
{4\pi G_4\over 3 \rho_0} {\rx^2 \over a^8}\;.
\label{3.691}
\ee
In this form, one can clearly see that the  problematic
terms from a   phenomenological viewpoint  - namely, the exotic stiff matter
($\sim a^{-6}$)
and the unorthodox quadratic  term ($\sim a^{-8}$)  - will rapidly
dilute with the expansion of the brane universe. 
Hence, this brane-world model remains quite plausible
in spite of the conspicuous deviations from the standard
picture.

An analytic solution for the above equation is not
generally obtainable. Nonetheless, our current interest
is in the  near-bounce solution, which is indeed
extractable by way of some appropriate simplifications.
More specifically, since the scale factor 
 attains its minimal  value at the bounce, 
we can safely disregard the constant vacuum ($\Lambda_4/3$)
and ``near-constant'' curvature ($k/a^2$) terms  relative to
the other contributors.
After defining
\be
A
\equiv {8\pi G_4\over 3}\left[\rx +{3\over 8\pi G_4}\omega M\right]\; ,
\label{3.693}
\ee
\be
B\equiv {\omega^2 M^2 \epsilon^2\over 4}
\label{3.694}
\ee
and
\be
C\equiv {4\pi G_4\over 3 \rho_0}\rx^2
\label{3.692}
\ee
(all of which are manifestly positive),
we are then  left with a comparatively  simple expression, 
\be
H^2=       
{A\over a^4}
-{B\over  a^6}+   {C \over a^8}\;.
\label{3.695}
\ee

With the assumption that a bounce does indeed
take place (that is, $H$ vanishes for some
finite value of  $a$),
the above equation is readily solvable, as recently demonstrated
in \cite{KTB}. 
Following the cited paper, we  first
introduce a pair of  new variables,
$x\equiv a^2$ and $dt\equiv a^2d\eta\;$, and then inspect the
resultant form,
\be
{1\over 4}\left(x^{\prime}\right)^2=Ax^2-Bx+C
\label{3.696}
\ee
(with a prime indicating a differentiation with respect to $\eta$).
The solution is then expressible as
\be
a^2=x={B\over 2A}+\sqrt{\Delta}\cosh\left[2\sqrt{A}\eta\right]
\label{3.697}
\ee
and
\be
t={B\over 2A}\eta+\sqrt{{\Delta\over 4A}}\sinh\left[2\sqrt{A}\eta\right]\;,
\label{3.698}
\ee
with
\be
\Delta\equiv\left({B\over 2A}\right)^2-{C\over A} >0\;.
\label{3.699}
\ee
Here, the (implied) integration constants have been conveniently fixed so
that $t$ and $\eta$ both vanish at the bounce;
that is,
when  $a^2=a^2_{min}\equiv {B\over 2A}+\sqrt{\Delta}\;$.

The inequality in the above equation (\ref{3.699})
is precisely what we obtained by constraining $H$
to vanish at finite $a$. This 
important  condition can also be expressed
as a lower bound on the charge ($\epsilon^2\sim Q^2$),
\be
\left({\omega M \epsilon\over 2}\right)^4> 2\alpha{\rx^3\over \rho_0} 
\left({8\pi G_4\over 3}\right)^2\;,
\label{3.6991}
\ee
where,
\be
\alpha\equiv 1+{3\omega M\over 8\pi G_4\rx }\;.
\label{ALP}
\ee
Note that  constraints from nucleosynthesis \cite{XXX,XX2}  
naturally limit the energy density of holographic radiation  
such that, at the very  most,   $\omega M/G_4 \sim {\cal O}[\rho_r]\;$;
 and so  $\alpha$ can  safely be regarded
as a constant of the order unity.

It should be emphasized that the above constraint (\ref{3.6991}),
along with the upper bound $\epsilon^2<1$ ({\it cf}, Eq.(\ref{3.4})),
severely restricts    the allowed range  of values for the  bulk charge.
This is somewhat different than the case of an
empty  ({\it modulo} induced matter) brane,
in which there is, {\it a priori}, only an upper bound on the charge.
That is, for an empty brane, a bounce will be obtained for 
any non-vanishing value of the charge, regardless of how
small \cite{MP,AJM}. (Note that  the difference between the
two scenarios can be  attributed to the presence,
or lack thereof, of a positive  term in the Friedmann equation
 that goes as $a^{-8}\;$.) 
It is, however, interesting  that, even
for an empty brane, holographic considerations
still imply the necessity for  a finite lower bound on the charge \cite{BBB}.
In the next section, we will find out what the
holographic principle  can tell us about the current 
(intrinsic matter) case.

\section{Causal Entropy Bound at the Bounce}

In this section, we will test the holographic viability
of our bounce cosmologies   by way of the
causal entropy bound \cite{CAU,BFV,BFM}.
Firstly, a brief explanation of this holographic
bound is in order. (For a  more detailed account as
relevant to the present context, see \cite{BBB}.)

The causal entropy bound can best be viewed
as a covariant generalization of its
predecessor, the Hubble entropy bound \cite{VEN} (also, \cite{EL1,EL2,EL3}).
The Hubble bound, itself, follows from
a pair of intuitive notions: 
{\it (i)} the entropy is maximized, in a given region
of space, by filling up the volume with maximal-sized black holes
and {\it (ii)} in a cosmological background, the maximal size that can
be achieved by a  
{\it stable}   black hole  is, roughly, dictated by the Hubble
horizon ({\it i.e.}, $R_S\sim H^{-1}$, where $R_S$ is the maximal
Schwarzschild radius). 

Given the above considerations and the black hole area law \cite{BEK,HAW}, 
the following bound can be deduced on the entropy, $S$,
contained in a region of volume $V$:  
\be
S <  {V\over H^{-3}} \times    {H^{-2}\over G_4} 
= {VH\over G_4}\;,
\label{2.4}
\ee
where a four-dimensional spacetime has been assumed and
  numerical factors of the order unity have been ignored. 

The premise of the causal entropy bound is
to replace the Hubble horizon with a ``causal
connection scale'', $R_{CC}$, that
can  be interpreted as the length scale above which
spacetime perturbations are causally disconnected.
(Presumably, a black hole could not maintain 
its stability over any greater distance than this.)
Although a highly technical process (see \cite{CAU} for
the gory details), the scale  $R_{CC}$  can, as it so happens, 
be expressed  in
an explicitly  covariant form. For our purposes,
however, it is sufficient to consider the expression as appropriate
for a spacelike slice of a FRW spacetime \cite{CAU}:
\be
R_{CC}^{-2}={\rm Max}\left[{\dot H}+2H^2+{k\over a^2},\quad  -{\dot H}+
{k\over a^2}\right]\;.
\label{2.5}
\ee
It is reassuring that, for a slowly evolving and flat spacetime,
one obtains the intuitive expectation,  $R_{CC}\sim H^{-1}\;$.

For future reference, 
the causal entropy  bound (in four dimensions) takes on the form  
\be
S< S_{CB}\equiv \beta {VR_{CC}^{-1}\over G_4}\;,
\label{2.9}
\ee
where $\beta$ is a ``fudge factor'' of the order unity
(reflecting any neglected constants and the inherent ambiguity in bounds
of this nature). There is significant  evidence
that $S_{CB}$ does, indeed, serve as  a true holographic bound
for physical spacetimes. (See \cite{CAU,BFV,BFM,BBB}  for
an elaboration.) Moreover, the formulation of the causal bound 
does  not assume, {\it a priori}, 
 any of the energy conditions of general
relativity.\footnote{For an elaboration on the energy conditions
in a brane-world context, see \cite{ZZZZZ}.} In this sense,
the causal  bound can play a  particularly useful role in
 the holographic discrimination of
exotic cosmologies.

As advertised, we will  now proceed  to test, via the above entropy
bound (\ref{2.9}), the exotic  bounce
cosmologies of Section  II.
Let us first point out that the causal bound
is known to  persist, automatically, for any spacetime that
contains  only non-exotic, causal  matter
at  non-Planckian temperatures \cite{BFM}. Now consider
that, for our bounce cosmologies in particular, both the
 exotic stiff matter and the (presumably acausal)  quadratic-density term 
are  rapidly 
diluted by the spacetime expansion. 
Hence,  for current purposes,
 it will be  sufficient to focus 
on the spacelike surface that describes the bounce 
($t=\eta=0$).

As an initial step, let us calculate  $R_{CC}$  by way
of Eq.(\ref{2.5}).
Since $H=0$ at the bounce and the curvature term
can be neglected, we have
\be
R_{CC}^{-2}=|{\dot H}| \quad\quad\quad {\rm at}\quad t=\eta=0\;.
\label{4.80}
\ee
We can calculate ${\dot H}$ by differentiating Eq.(\ref{3.695}).
Also employing Eq.(\ref{3.697}) for $a^2$ and Eq.(\ref{3.699}) for
$\Delta$, we
obtain
\be
R_{CC}^{-2}={A\over a^4}\left|1-{C\over Aa^4}\right|
\quad\quad\quad {\rm at} \quad t=\eta=0\;.
\label{4.81}
\ee

It will  prove to be convenient if
the scale factor is re-expressed  in terms of the ``total'' 
energy density of radiation,
\be
\rho_R \equiv {1\over a ^4}\left[\rx + {3\over 8\pi G_4}\omega M\right]\;,
\label{4.82}
\ee
from  which it follows that ({\it cf},  Eqs.(\ref{3.693},\ref{ALP})) 
\be
\rho_R  ={\alpha\over a^4}\rx
= {3\over 8\pi G_4}{A\over a^4}\;,
\label{4.83}
\ee
and so,
\be
R_{CC}^{-2}={8\pi G_4\over 3}\rho_R
\left[1-{1\over 2 \alpha^2 }{\rho_R\over \rho_0}\right]
\quad\quad\quad {\rm at} \quad t=\eta=0\;, 
\label{4.84}
\ee
where we have also applied Eq.(\ref{3.692}).

 Next, we will  evaluate the  entropy contained in this
spacelike bounce surface.  As the  entropy of radiation
can be expected to dominate over any other  type of matter,
it should  be  sufficient to consider the contribution from
just the radiative sources. To further simplify matters, let
us  assume that the effective number
of particle species is roughly  the same for both the
standard-model  and holographic forms of radiation.
(As  can be seen in the subsequent analysis, any
 discrepancy of $10^{4}$ or smaller  
is inconsequential up to factors of the order
unity.) In this way, we can  define both a
{\it total} entropy, $S_R$, and entropy density, $s_R$,
 and  be able to  relate these
to   $\rho_R$ (\ref{4.82}) in a straightforward manner.     

It is helpful, at this stage,  to recall 
the  Stephan-Boltzmann thermodynamic relations
for radiative matter (in thermal equilibrium at temperature $T$):
\be
s_R = {S_R\over V} = {\cal N}  T^3\;,
\label{4.5}
\ee
\be
\rho_R ={\cal N}  T^4\;,
\label{4.6}
\ee
where the missing numerical factors (obviously of
the order unity) have been   absorbed
into the effective number of particle species, ${\cal N}$.
 Eliminating $T$ from this pair of equations,
we find that
\be
S_R=V{\cal N}^{1/ 4} \rho_R^{3/ 4}\;.
\label{4.7}
\ee

Substituting the above outcome into
the left-hand side of Eq.(\ref{2.9}) and our prior
result for $R_{CC}$ (\ref{4.84}) into
the right-hand side, we  are able to establish
 the following inequality (up to overall factors of
the order unity\footnote{Here, we are assuming
that ${\cal N}$ is not significantly greater than $10^4$, which would
seem to be a reasonable enough constraint. This    number also agrees, 
coincidentally, with an
upper limit that ensures the stability of
a  Minkowski vacuum \cite{WHY}.})
\be
G_4 \sqrt{\rho_R} <\left|1-{1\over 2 \alpha^2 }{\rho_R\over \rho_0}\right|
\quad\quad\quad {\rm at} \quad t=\eta=0\;.
\label{4.85}
\ee

Hence, we have obtained, by virtue of the holographic principle, 
an upper bound on the energy density of 
radiation at the bounce.  What may not be immediately clear
is the existence of  a lower bound as well. To elaborate,
if we regard the unscaled charge ($Q^2$) as a fixed quantity, then
$M$ (and, hence, the total radiation density via  Eq.(\ref{4.82})) 
must remain sufficiently large to ensure that
the dimensionless charge ($\epsilon^2$)
is always less than unity; {\it cf}, Eq.(\ref{3.4}).
Moreover, phenomenologically speaking, any lower
bound  on $M$ will translate directly into a lower bound
on $\rho_r$ ({\it i.e.}, the energy density of standard-model radiation);
inasmuch as nucleosynthesis considerations must limit 
the relative contribution of the holographic radiation
 \cite{XXX,XX2}.
(Alternatively,  $\rho_R=\alpha\rho_r\;$,
where compliance with observations suggests that $\alpha\sim{\cal O}[1]\;$.)
Putting this altogether, we see that holographic considerations 
help to significantly constrain the allowable values of $\rho_r$.

To get a better feel for the derived bound (\ref{4.85}),
it is  instructive  to consider 
 the ``natural'' limiting case, 
$\rho_R/\rho_0<<1\;$. (That is to say, it is quite natural to take, at 
the very least, $\rho_R/\rho_0<1\;$;
as this ensures  that the quadratic energy-density term is subdominant
to the linear term at the time of nucleosynthesis \cite{MP,KTB}.)
In this limit, the situation essentially
simplifies to the one studied in our prior paper \cite{BBB}
(except that the model in \cite{BBB} contained
purely holographic radiation)  and one can show that
\be
\epsilon^2 >
{G_4^2 \left[\alpha\rx\right]^{3/2}\over \omega^2 M^2}
>\left[\alpha\rx\right]^{-1/2} \;.  
\label{4.86}
\ee
(As usual, up to numerical factors  and also  note that  the second inequality 
assumes 
  the  induced radiation  to be the subdominant source.)
Which is to say, the  upper bound on the energy
density of radiation can also be interpreted  as  a lower bound on
the bulk  charge. Interestingly, a lower bound
on the charge is also necessitated by the
existence of a bounce; {\it cf}, Eq.(\ref{3.6991}).

The deep  implications of  the holographic principle on these brane worlds 
 can be
further illuminated  in the following way.
 First of all, let us  rewrite our  prior
inequality (\ref{4.85}) in terms of
the parameters $A$ and $C$  (\ref{3.693},\ref{3.692}):
\be
G_4 A<\left[{a^4A-C\over a^2 A}\right]^2
\quad\quad\quad {\rm at} \quad t=\eta=0\;,
\label{4.87}
\ee
where we have also utilized Eq.(\ref{4.83}) for
$\rho_R\;$,  and note that, although {\it overall} numerical factors 
have been neglected,
 the expression inside
the square brackets is exact.

Next, we can apply the relation for $a^2$ at the bounce
({\it cf}, Eq.(\ref{3.697})) to obtain (up to
overall factors)
\be
G_4 A<{1\over A^2}\left[{\left({B\over 2}\right)^2+ AB\sqrt{\Delta}+
A^2\Delta- AC\over B+2A\sqrt{\Delta}}\right]^2\; .
\label{zzz}
\ee

Finally, the definition of $\Delta$ (\ref{3.699})
(and some simplification)
can be employed to yield
a  remarkably concise form,
\be
G_4 A< \Delta \;. 
\label{4.89}
\ee
Since $A$ is a manifestly positive quantity,
one has, as an immediate consequence of the
holographic paradigm, that $\Delta$ must 
be strictly positive. However, this
is just the constraint (\ref{3.699}) that we  previously  
{\it assumed} so as to
ensure that a bounce does indeed take place.
That is to say,  when a brane is moving in the background
of a charged 
black hole, then compliance with the causal entropy bound
is a sufficient prerequisite for the existence of a bounce.

\section{Conclusion}

To summarize, we have been considering the
 implications of  holography  on  a certain class
of brane-world models; more specifically, a four-dimensional
brane world moving in the five-dimensional anti-de Sitter  background
of a charged black hole.  Such models are of significant
interest because they allow for the possibility
of a non-singular bounce (as opposed to a big bang/crunch),
although at the expense of (holographically induced) exotic matter. 

In a previous related paper \cite{BBB},
such cosmologies were studied, from a holographic perspective,
for the very special case of a  brane that
is void of any intrinsic  matter sources.
 Meanwhile, in the current treatment, we
took a step in the direction of realism and generalized 
considerations to a brane that does indeed contain
standard-model matter.  In particular,
radiative matter was incorporated,
as this form complies with empirical expectations
(for the small-scale universe) and provides the most 
challenging test for any holographic bound.

For the purposes of testing holographic viability, we
called upon  the so-called causal
entropy bound \cite{CAU}. Significantly,  this  holographic bound
does not assume, {\it a priori},  any of the usual energy conditions,
and so is particularly 
well suited for the discrimination of exotic cosmologies.
Ultimately, we found that the causal  bound implies
a lower limit on  the (bulk) black hole
charge or, equivalently, an upper limit on the  energy
density  of the radiative matter (including both standard-model and 
holographically induced contributions). Moreover, we have demonstrated that
compliance with the causal bound is a sufficient condition
for a brane universe to avoid a singular collapse.

Finally, 
it is interesting to recall our earlier related study \cite{BBB} in light
of these new findings.
Even when the brane is void of intrinsic matter, 
the causal bound implies 
a finite lower limit on the magnitude  of the bulk charge.
However, somewhat paradoxically, a bounce is always
realized in  this (empty-brane) class of models.
 That is to say, the ``motivation'' for
a lower bound on the bulk charge seems to
be missing in the empty-brane scenario. 
It may be of interest to see if similar patterns 
of behavior show up in other types of
bounce cosmologies.

\section{Acknowledgments}
\par
The author  would like to thank  V.P.  Frolov  for helpful
conversations and R. Bousso for  constructively criticizing
the  original version of the paper.



\end{document}